\documentstyle[aps,prl,psfig,epsfig,twocolumn,psfrag,amsfonts]{revtex}
\input{epsf}
\draft
\begin{document}
\title{Fractal structures of normal and anomalous diffusion\\ 
in nonlinear nonhyperbolic dynamical systems}
\author{N. Korabel\cite{emkor} and R. Klages\cite{emkla}}
\address{Max-Planck-Institut f\"ur Physik komplexer Systeme, N\"othnitzer
Strasse 38, D-01187 Dresden, Germany}
\date{\today}
\wideabs{
\maketitle
\begin{abstract}
A paradigmatic nonhyperbolic dynamical system exhibiting deterministic
diffusion is the smooth nonlinear climbing sine map. We find that this map
generates fractal hierarchies of normal and anomalous diffusive regions as
functions of the control parameter. The measure of these self-similar sets is
positive, parameter-dependent, and in case of normal diffusion it shows a
fractal diffusion coefficient. By using a Green-Kubo formula we link these
fractal structures to the nonlinear microscopic dynamics in terms of fractal
Takagi-like functions.
\end{abstract}
}

It is well-known that diffusion processes in nonlinear dynamical systems may
be generated by microscopic deterministic chaos in the equations of
motion. This fact points to a somewhat deeper foundation of nonequilibrium
statistical mechanics than modeling diffusive transport by stochastic random
walks. In order to understand deterministic diffusion, statistical mechanics
was suitably combined with dynamical systems theory \cite{Lich,Ott,Gas}. Much
was learned by analyzing simple models of deterministic transport such as one-
and two-dimensional maps \cite{Gei82,SchGro,Dan89,Rec80}, chaotic billiards in
external fields \cite{Har01}, periodic Lorentz gases
\cite{KlaDel,Kla02}, and certain differential equations \cite{Chi79}. However,
characteristics of deterministic diffusion were also observed experimentally,
that is, in dissipative systems driven by periodic forces such as Josephson
junctions in the presence of microwave radiation \cite{JJ}, in superionic
conductors \cite{SupCond}, and in systems exhibiting charge-density waves
\cite{CDW}. The equations of motion of these systems are generally of the form
of some nonlinear pendulum equation. In the limiting case of strong
dissipation, these differential equations were reduced to nonhyperbolic
one-dimensional maps sharing certain symmetries \cite{Bohr}. The so-called
climbing sine map is a well-known example of this class of maps
\cite{Gei82,SchGro}. Due to its nonhyperbolicity, the map posseses a rich
dynamics consisting of chaotic diffusive motion, ballistic dynamics, and
localized orbits. Under parameter variation these different types of dynamics
are highly intertwined resulting in complicated scenarios related to the
appearance of periodic windows \cite{SchGro}. On the other hand, for simple
one-dimensional hyperbolic maps it was shown that the diffusion coefficient is
typically a fractal function of control parameters \cite{Kla95,Kla}. An
analogous behavior was also detected for other transport coefficients
\cite{GaKl}, and in more complicated models \cite{Har01,KlaDel,Kla02}. 
However, up to now the fractality of transport coefficients could be assessed
for hyperbolic systems only, whereas, to our knowledge, the fractal nature of
classical transport coefficients in the broad class of nonhyperbolic systems
was not discussed.

Here we show that the nonhyperbolicity of the climbing sine map does not
destroy these fractal characteristics of deterministic diffusive transport. On
the contrary, fractal structures appear for normal diffusive parameters as
well as for anomalous diffusive regions. We argue that higher-order memory
effects are crucial to understand the origin of these fractal hierarchies in
nonhyperbolic systems. By using a Green-Kubo formula for diffusion, the
dynamical correlations are recovered in terms of fractal Takagi-like
functions. We furthermore show that the distribution of periodic windows forms
devil's staircase-like structures as a function of the parameter and that the
complementary sets of chaotic dynamics have a positive measure in parameter
space that increases by increasing the parameter value.

The climbing sine map we study is defined as
\begin{equation}
x_{n+1} = M_a(x_n) \; , \; \; M_a (x) := x + a \sin(2 \pi x)\:,
\label{model}
\end{equation}
where $a\in\mathbb{R}$ is a control parameter, $x\in\mathbb{R}$, and $x_n$ is
the position of a point particle at discrete time $n$. Obviously, $M_a(x)$
possesses translation and reflection symmetry,
\begin{equation}
M_a(x + m) = M_a (x) + m \;, \; \; M_a(-x) = -M_a (x)\:.
\label{sym}
\end{equation}
The periodicity of the map naturally splits the phase space into different
cells $(m, m+1]$, $m \in \mathbb{Z}$. We will be interested in parameters $a >
0.732644$ for which the extrema of the map exceed the boundaries of each
cell for the first time indicating the onset of diffusive motion.

\begin{figure}[htb]
\psfrag{x}{$\tilde{x}$}
\psfrag{b}{$a$}
\psfrag{D_b}{$D_a$}
\vspace{20pt}
\centerline{\psfig{figure=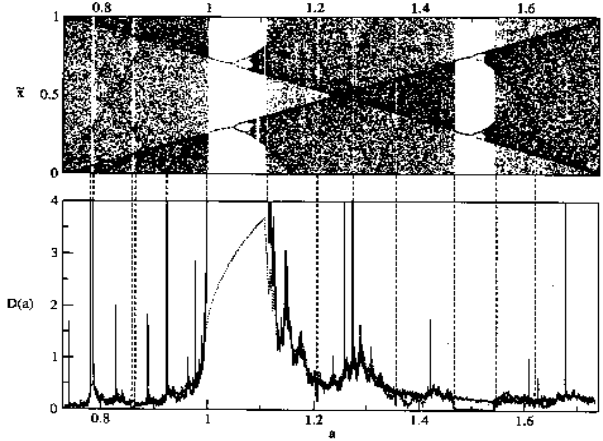,width=82mm,height=60mm}}
\vspace{2pt}
\caption{Upper panel: bifurcation diagram for the climbing sine map.
Lower panel: diffusion coefficient computed from simulations as a function of
the control parameter $a$ in comparison with the correlated random walk
approximation $D_{10}^1(a)$ (dots). The dashed vertical lines connect regions
of anomalous diffusion, $D(a)\to\infty$ or $D(a) \to 0$, with ballistic and
localized dynamics in respective windows of the bifurcation diagram.}
\label{Dif}
\end{figure}

The bifurcation diagram of the associated reduced map $\tilde{M}_a (\tilde{x})
:= M_a (x) \; \mbox{mod} \; 1\:,\:\tilde{x} := x \; mod \; 1$, consists of
infinitely many periodic windows, see Fig.\ \ref{Dif}. Whenever there is a
window the dynamics of Eq.\ (\ref{model}) is either ballistic or localized
\cite{SchGro}. Fig.\ \ref{Dif} demonstrates that this scenario has a strong impact on
the diffusion coefficient defined by $D(a ):=\lim_{n \rightarrow \infty}
\langle x_n^2 \rangle /(2n)$, where the brackets denote an ensemble average
over moving particles. For localized dynamics orbits are confined within some
finite interval in phase space implying subdiffusive behavior for which the
diffusion coefficient vanishes, whereas for ballistic motion particles
propagate superdiffusively with the diffusion coefficient being proportional
to $n$. Only for normal diffusion $D(a)$ is nonzero and finite.  At the
boundaries of each periodic window there is transient intermittent-like
behavior eventually resulting in normal diffusion with $D(a)\sim a^{(\pm
\frac{1}{2})}$ \cite{Gei82,SchGro}. Here we are interested in the complete
parameter-dependent diffusion coefficient. For this purpose we compute $D(a)$
from numerical simulations by using the Green-Kubo formula for maps
\cite{Gas,Kla02,Kla95,GaKl},
  
\vspace*{-1cm}
%
%
  
\vspace*{-0.5cm}
\noindent

\begin{equation}
D_n(a) = \langle j_a(\tilde{x}_0) J_a^n(\tilde{x}) \rangle -
\frac{1}{2}\langle j_a^{2}(\tilde{x}_0) \rangle \:,
\label{GK}
\end{equation}
where the angular brackets denote an average over the invariant density of the
reduced map, $\langle \ldots \rangle := \int d\tilde{x}\rho
(\tilde{x})\ldots$. The jump velocity $j_a$ is defined by
$j_a(\tilde{x}_n):=[x_{n+1}]-[x_n]\equiv [M_a(\tilde{x}_n)]$, where the square
brackets denote the largest integer less than the argument.  The sum $J_a^n
(\tilde{x}) := \sum_{k=0}^n j_a (\tilde{x}_k)$ gives the integer value of the
displacement of a particle after $n$ time steps that started at some initial
position $x\equiv x_0$, and we call it jump velocity function. Eq.\ (\ref{GK})
defines a time-dependent diffusion coefficient which, in case of normal
diffusion, converges to $D(a)\equiv\lim_{n\to\infty}D_n(a)$. In our
simulations we truncated $J_a^n(\tilde{x})$ after having obtained enough
convergence for $D(a)$, that is, after $20$ time steps. The invariant density
was obtained by solving the continuity equation for $\rho (\tilde{x})$ with
the histogram method of Ref.\
\cite{Lich}.

The highly non-trivial behavior of the diffusion coefficient in Fig.\
\ref{Dif} can qualitatively be understood as follows: The
Green-Kubo formula Eq.\ (\ref{GK}) splits the dynamics into an inter-cell
dynamics, in terms of integer jumps, and into an intra-cell dynamics, as
represented by the invariant density. We first approximate the invariant
density in Eq.\ (\ref{GK}) to $\rho (\tilde{x})\simeq 1$ irrespective of the
fact that it is a complicated function of $x$ and $a$ \cite{SchGro}. This
approximate diffusion coefficient we denote with a superscript in Eq.\
(\ref{GK}), $D_n^1(a)$. The term for $n=0$ is well-known as the stochastic
random walk approximation for maps, which excludes any higher-order
correlations \cite{Gei82,SchGro,Kla}. The generalization $D_n^1(a)\;,\;n>0$
was called correlated random walk approximation
\cite{Kla02}. We now use this systematic expansion to analyze the
diffusion coefficient of the climbing sine map in terms of higher-order
correlations.

\vspace*{-1cm}
%
%
\begin{figure}[htb]
\psfrag{D^n_b}{$D_a^n$}
\psfrag{b}{$a$}
\psfrag{(a)}{$(a)$}
\psfrag{(b)}{$(b)$}
\psfrag{(c)}{$(c)$}
\vspace{20pt}
\centerline{\psfig{figure=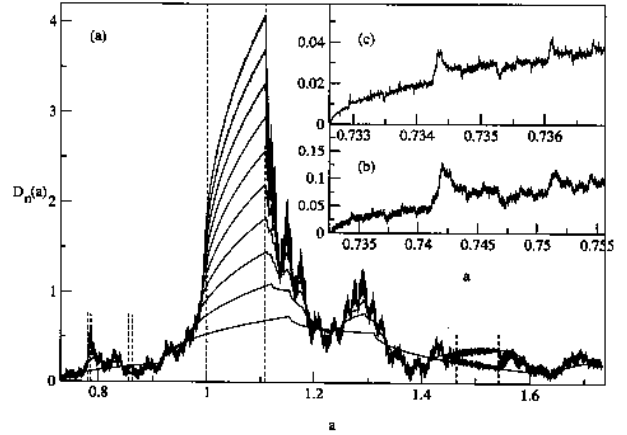,width=82mm,height=60mm}}
\vspace{2pt}
\caption{(a) Sequence of correlated random walks $D_n^1(a)$ for $n=1, \cdots,
10$. Note the quick convergence for normal diffusive parameters. The dashed
lines define the same periodic windows as in Fig.\ref{Dif}. The inserts (b)
and (c) contain blowups of $D_{10}^1(a)$ in the initial region of (a). They
show self-similar behavior on smaller and smaller scales.}
\label{RW}
\end{figure}
\noindent

In Fig.\ref{RW} (a) we depict results for $D_n(a)$ at $n=1,\ldots,10$. One
clearly observes convergence of this approximation in parameter regions with
normal diffusion. Indeed, a comparison of $D_{10}^1(a)$ with $D(a)$, as shown
in Fig.\ref{Dif}, demonstrates that there is qualitative agreement on large
scales. On the other hand, for parameters corresponding to ballistic motion
the sequence of $D_n^1(a)$ diverges, in agreement with $D(a)\to\infty$,
whereas for localized dynamics it alternates between two solutions. This
oscillation is reminiscent of the dynamical origin of localization in terms of
certain period-two orbits. That these solutions are non-zero is due to the
fact that the invariant density was approximated. In regions of normal
diffusion this approximation nicely reproduces the irregularities in the
diffusion coefficient. Even more importantly, the magnifications in
Fig.\ref{RW} give clear evidence for a self-similar structure of the diffusion
coefficient. Our results thus show that dealing with correlated jumps only
yields a qualitative understanding of normal and anomalous diffusion in the
climbing sine map.

We now further analyze the dynamical origin of these different structures.
According to its definition, the time-dependent jump velocity function $J_a^n
(\tilde{x})$ fulfills the recursion relation
\begin{equation}
J_a^n (\tilde{x}) =  j_a(\tilde{x}) + J_a^{n-1} (M_a(\tilde{x})) \;.
\label{rec}
\end{equation}
$J_a^n(\tilde{x})$ is getting extremely complicated after some time steps,
thus we introduce the more well-behaved function
\begin{equation}
T_a^n(\tilde{x}) := \int_0^x J_a^n (z) \; dz, \; \; \; T_a^n(0)\equiv
T_a^n(1) \equiv 0.
\label{Takagi}
\end{equation}
  
\vspace*{-1.4cm}
%
%
\begin{figure}[htb]
\psfrag{T}{$T_a^n$}
\psfrag{x}{$\tilde{x}$}
\psfrag{a}{$(a)$}
\psfrag{b}{$(b)$}
\vspace{20pt}
\centerline{\psfig{figure=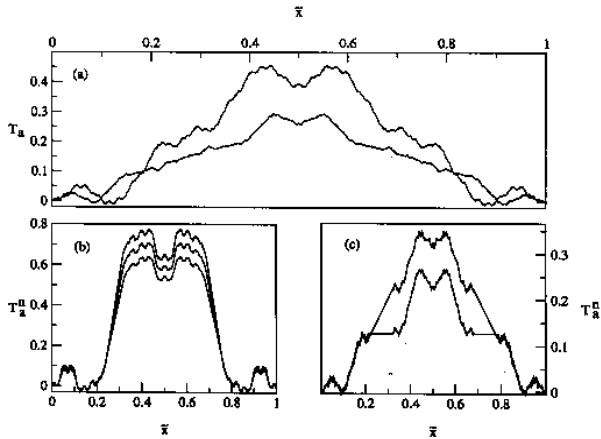,width=82mm,height=60mm}}
\vspace{2pt}
\caption{Functions $T_a^n(\tilde{x})$ for the climbing sine map representing
the number of jumps particles starting at $\tilde{x}$ have traveled when this
number is integrated from zero to $\tilde{x}$, see Eqs.\
(\ref{rec})-(\ref{TakEq2}).  (a) depicts diffusive dynamics at $a=1.2397$
(upper curve) and at $a=1.7427$ (lower curve), (b) ballistic dynamics at
$a=1.0$, and (c) localized dynamics at $a=1.5$. In (a) the limiting case for
$n\to\infty$ is shown, in (b) and (c) it was $n=5, 6, 7$. Note the divergence
in (b) and the oscillation in (c).}
\label{TakFIG}
\end{figure}
%
\noindent 
Integration of Eq.\ (\ref{rec}) thus yields the recursive functional equation
\begin{equation}
T_a^n(\tilde{x}) = t_a (\tilde{x}) + \frac{1}{\tilde{M}_a(\tilde{x})} \;
T_a^{n-1} (\tilde{M}_a(\tilde{x})) - I(\tilde{x})
\label{TakEq}
\end{equation}
with the integral term
\begin{equation}
I(\tilde{x}) := \int_{0}^{\tilde{M}_a(\tilde{x})}\; dz g''(z) T_a^{n-1} (z)\;,
\label{TakEq2}
\end{equation}
where $t_a (\tilde{x}) := \int dz\; j_a (z)$, and $g^{''}(z)$ is the second
derivative of the inverse function of $\tilde{M}_a (\tilde{x})$
\cite{note1}. For piecewise linear hyperbolic maps $I(\tilde{x})$ simply
disappears and the derivative in front of the second term reduces to the local
slope of the map thus recovering ordinary de Rham-type equations
\cite{Gas,Kla,GaKl}.  It is not known to us how to directly solve this generalized
de Rham-equation for the climbing sine map, however, solutions can
alternatively be constructed from Eq.\ (\ref{Takagi}) on the basis of
simulations. Results are shown in Fig.\ \ref{TakFIG}.  For normal diffusive
parameters the limit $T_a(\tilde{x})=\lim_{n \to\infty} T_a^n(\tilde{x})$
exists, and the respective curve is fractal over the whole unit interval
somewhat resembling (generalized) fractal Takagi functions
\cite{Gas,Kla,GaKl}. However, in case of periodic windows $T_a^n(\tilde{x})$
either diverges due to ballistic flights, or it oscillates indicating
localization. Interestingly, in these functions the corresponding attracting
sets appear in form of smooth, non-fractal regions on fine scales, whereas the
other regions appear to be fractal.

The diffusion coefficient can now be formulated in terms of these fractal
functions by integrating Eq.\ (\ref{GK}). For $a\in(0.732644,1.742726]$ we get
\begin{equation}
D(a)= 2\left[T_a(\tilde{x}_2)\rho (\tilde{x}_2)-T_a(\tilde{x}_1)\rho
(\tilde{x}_1)\right] - D_0^{\rho}(a),
\label{DTak}
\end{equation}
where $\tilde{x}_i,\;i=1,2$, is defined by $[M_a(\tilde{x}_i)]:=1$, and
$D_0^{\rho}(a):=\int_{\tilde{x}_1}^{\tilde{x}_2} d\tilde{x} \rho
(\tilde{x})$. Our previous approximation $D_n^1(a)$ is recovered from this
equation as a special case.

The intimate relation between periodic windows and the irregular behavior of
the diffusion coefficient motivates us to investigate the structure of the
periodic windows in the climbing sine map in more detail. The appearance of
windows was analyzed quite extensively for non-diffusive unimodal maps
\cite{Win}, whereas for diffusive maps on the line, apart from the preliminary 
studies of Refs.\ \cite{SchGro}, nothing appears to be known. The windows are
generated by certain periodic orbits, consequently there are infinitely many
of them, and they are believed to be dense in the parameter set
\cite{Ott}. Windows with ballistic dynamics are born through tangent
bifurcations, further undergo Feigenbaum-type scenarios and eventually
terminate at crisis points. Windows with localized orbits only occur at even
periods. They start with tangent bifurcations and exhibit a symmetry breaking
at slope-type bifurcation points.
  
\vspace*{-0.8cm}
%
\begin{figure}[b]
\psfrag{N}{$N$}
\psfrag{b}{$a$}
\vspace{20pt}
\centerline{\psfig{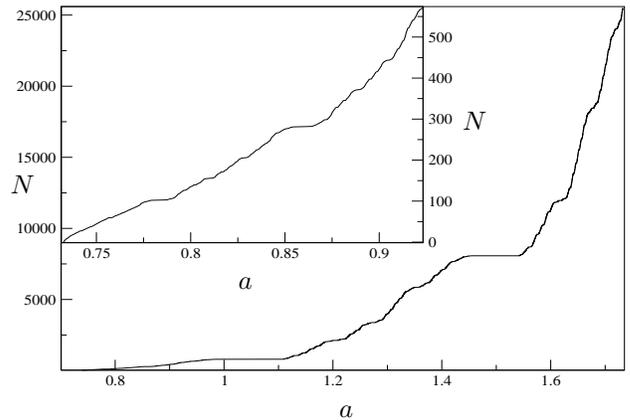}}
\vspace{2pt}
\caption{Devil's staircase-like structure formed by the distribution of
periodic windows as a function of the control parameter.  $N$ is the
integrated number of period six-windows.  The inset shows a blowup of the
initial region.}
\label{Anomal}
\end{figure}
\noindent

In order to analyze the structure of the regions of anomalous diffusion, we
sum up the number of period six-windows as a function of the parameter, that
is, the total number is increased by one for any parameter value at which a
new period six-window appears. This sum forms a devil's staircase-like
structure in parameter space indicating an underlying Cantor set-like
distribution for the corresponding anomalous diffusive region, see Fig.\
\ref{Anomal}. The (Lebesque) measure of periodic windows is obviously
positive, hence this set must be a fat fractal \cite{Yor}. Its
self-similar structure can quantitatively be assessed by computing the
so-called fatness exponent \cite{Far}. We are furthermore interested in the
parameter dependence of this fractal structure, therefore we divide the
parameter line into subsets labeled by the integer value of the map maximum on
the unit interval, $[M_a(x_{max})]=j$, $j \in \mathbb{Z}$. For $j=1,2,3$ we
obtain a fatness exponent of $0.45$ with errors of $0.03$, $0.04$, and $0.05$
for the different $j$. We mention that this value was conjectured to be
universal and was also obtained for non-diffusive unimodal maps
\cite{Far}.

We now study the measure of the windows as a function of the parameter. For
this purpose we computed all windows up to period six for the first subset, up
to period $5$ for $j=2,3$, and we summed up their measures in the respective
subsets. We find that the total measure decays exponentially as a function of
$j$ while oscillating with odd and even values of $j$ on a finer scale
\cite{Kor}. This oscillation can be traced back to windows generated by
localized dynamics, which only appear at even periods thus contributing only
periodically to the total measure. However, different measures of `ballistic`
and `localized` windows decay with the same rate. We have furthermore computed
the complementary measure $C_j$ of diffusive dynamics in the $j$th subset of
parameters.  We find that $C_1=0.783$, $C_2=0.808$, and $C_3=0.932$ with an
error of $\pm 0.002$, so the measure of the diffusive regions is always
non-zero and seems to approach one with increasing parameter values.

We finally remark that the climbing sine map is of the same functional form as
the respective nonlinear equation in the two-dimensional standard map, which
is considered to be a standard model for many physical, Hamiltonian dynamical
systems. Indeed, both models are motivated by the driven pendulum, both are
strongly nonhyperbolic, and although the standard map is area-preserving it
too exhibits a highly irregular parameter-dependent diffusion
coefficient. Understanding the origin of these irregularities was the subject
of intensive research \cite{Ott,Rec80}, however, so far the complexity of the
system prevented to reveal its possibly fractal nature. A suitably adapted
version of our approach to nonhyperbolic diffusive dynamics may enable to make
some progress in this direction.



\begin{thebibliography}{10}
\bibitem{note1}
In order to define an inverse of $\tilde{M}(\tilde{x})$, we split the unit
interval into subintervals on which this function is piecewise invertible.
\end{thebibliography}
\end{document}